\documentclass[a4paper, 12pt]{article}
\usepackage{epsfig}
\begin{document} 


\title{On the absence of conduction electrons in the
antiferromagnetic part of the phase-separated
states in magnetic semiconductors
\footnote{Published in \textit{Physics Letters A}, \textbf{298} (2002),
p. 185-192}
}

\author{I. Gonz\'alez\thanks{E-mail:faivan@usc.es}, J. Castro, D.
Baldomir\\
\small{\textit{Departamento de F\'{\i}sica Aplicada,}}\\
\small{\textit{Universidade de Santiago de Compostela,}} \\
\small{\textit{E-15706 Santiago de Compostela, Spain.}}
}

\maketitle

\begin{abstract}
We have calculated the energies of the phase-separated states for      %
degenerate antiferromagnetic semiconductors including the 
possibility of the existence of conduction electrons in the
antiferromagnetic part of the phase-separated states. It is
demonstrated that, at $T=0$, the minimum energy corresponds to a 
droplet phase with absence of electrons in the antiferromagnetic part. %
\end{abstract}


\section{Introduction}

Degenerate antiferromagnetic semiconductors are obtained by strongly   %
doping antiferromagnetic semiconductors (e.g., $EuTe$ or lanthanium
manganites). Over a concentration range of doping impurities, the
ground state of these compounds will be a mixture of antiferromagnetic
(AF) and ferromagnetic (FM) phases \cite{Nag83,Nag72,Kas73}. Starting 
from a pure compound and doping, it is found that the ground state is 
AF and non-conducting up to some conduction electron concentration,
driven by impurities, $n_{\mathrm{d}}$, for which the homogeneous state 
turns to be unstable against phase separation. The ground state becomes 
then inhomogeneous with a simply connected AF phase and a multiply
connected FM phase. The exact geometry of the multiply connected phase 
is of fractal nature. On increasing the doping, a geometric transition 
takes place in which the topology of the sample changes due to
percolation of the FM phase. Now, the ground state corresponds to a 
multiply connected AF phase and a simply connected FM one. The doping 
concentration at which this transition occurs is  denoted by 
$n_{\mathrm{T}}$. At this point an important change in the conductivity
is expected, because the material becomes a conductor. On a further 
increase of doping, a certain concentration, $n_{\mathrm{u}}$, exists 
at which this phase-separated state starts to be unstable and the 
sample becomes again homogeneous but now a FM. Transitions between 
these phases are also observed by varying magnetic field and 
temperature \cite{Nag90}. 

The physical reason for having a phase-separated state in an           %
antiferro\-mag\-netic semiconductor is the dependence of the energy of 
the charge carriers that appear on doping on the magnetic order of the
lattice. The charge carrier energy is lower if they move in a FM 
region. Therefore, by interaction with the spin system in the lattice, 
they are able to change its magnetic order, creating FM micro-regions
and become trapped in them. In a degenerate magnetic semiconductor 
this is a cooperative phenomenon: a number of carriers are 
self-trapped in the same FM micro-region reducing the energy per
carrier necessary to create it. When the carrier concentration is high
enough, the phase-separated state turns out to be stable.
It must be stressed that our calculation is valid only in the case of 
a degenerate semiconductor, i.e. over the range of conduction electron
density in which this collective behaviour of the carriers appears. 
The micro-region contains a high number of conduction electrons.
No individual (isolated) ferrons (magnetic polarons) exists over this
range of concentrations. Unfortunately, there is no means of 
calculate $n_{\mathrm{d}}$ accurately, simply it must be defined as the 
density from which these micro-regions contain a ``high'' number of 
conduction electrons. 
   
Within the Vonsovsky s-d model and using a variational procedure to    %
calculate the energies of the different phases, Nagaev shows that the
phase-separated state was the ground  state for degenerate
antiferromagnetic semiconductors over a certain range of doping. One 
of the assumptions of the calculation is the absence of conduction
electrons in the AF part of the phase-separated state (see the
reference \cite{Nag83} for a review). In this article we modifed the
calculations in \cite{Nag83} to allow for the presence of conduction 
electrons in the AF part of phase-separated state and we shown that the 
energy minimum is encountered when all the conduction electrons are in
the FM phase.

\section{The Hamiltonian of the system} 

Our starting point is the Hamiltonian of the generalized Vonsovsky     %
s-d model:

\begin{eqnarray}\label{ hamiltonian}
H & = & \sum E_{\vec{k}}a^{\dagger}_{\vec{k}\sigma}a_{\vec{k}\sigma}-
\frac{A}{N}\sum \vec{S}_{\vec{h}}\vec{s}_{\sigma \sigma'}
\exp[i(\vec{k}-\vec{k'})\vec{g}] \nonumber\\
& & \times a^{\dagger}_{\vec{k}\sigma}a_{\vec{k'}\sigma'}-\frac{1}{2}
\sum I(\vec{g}-\vec{f})\vec{S}_{\vec{g}}\vec{S}_{\vec{f}} + H_{C}
\end{eqnarray} 
where $a^{\dagger}_{\vec{k}\sigma},\,a_{\vec{k}\sigma}$ are the        %
creation and destruction operators of a conduction electron with 
quasi-momentum $\vec{k}$ and spin projection $\sigma$, 
$\vec{S}_{\vec{g}}$ is the operator of the core spins of the magnetic 
ions (d-spins) with number $\vec{g}$, $\vec{s}_{\sigma \sigma'}$ are 
the Pauli matrices and $N$ is the total number of unit cells in the 
crystal. The crystal structure is assumed to be simple cubic, the 
d-spin magnitude being S. In the case of that the conduction electrons 
move on the magnetic ions the label $\vec{h}$ in the second term of 
equation (\ref{ hamiltonian}) is equal to $\vec{g}$. But in the case
of the electron moving on a non-magnetic atom then $\vec{h}$ 
corresponds to the nearest magnetic neighbors of this ion and 
summation is carried over them. The term bilinear in the d-spin
operators will be called the exchange 
Hamiltonian, though in reality it corresponds to the super-exchange
interaction between magnetic ions. The term $H_{C}$ describes the
Coulomb energy of interaction of the conduction electrons with each
other and with ionized impurities. The charge of the latter,
compensating the charge of the conduction electrons, is assumed to be 
distributed uniformly over the sample (the jellium model).

The main parameters in the Vonsovsky s-d model are $W=2zt$,            %
the carriers band-width, $AS$ the exchange energy between conduction
electrons and magnetic ions (s-d exchange energy) and $zIS^2$ the 
exchange energy between magnetic ions (d-d exchange energy). $I<0$ is
the exchange integral between first-nearest neighbors magnetic ions.
The smallest parameter is the d-d exchange energy, which is of the 
order of magnitude of N\'eel temperature of the pure compound (without 
doping). We differentiate two possibilities depending on the relative 
value of $W$ and $AS$. In the case $W>>AS$ we have a wide-band 
semiconductor (e.g. $EuTe$). In the opposite case $W<<AS$, we talk 
about a ``double-exchange'' material (e.g. lanthanium 
manganites). For the sake of definiteness the sign of $A$ is assumed
to be positive. We also consider only the case $H=0$ because at least
up to fields $H\leq2J$ (the field at which both sublattices collapse)
the difference in energy for FM and AF parts always favored the
conduction electron being in the FM part.

\section{Variational procedure}

As in references \cite{Nag83,Nag72,Kas73} a variational procedure      %
will be used here to find the ground state energy of the 
phase-separated system. Three variational parameters are used, the 
ratio $x$ of the volumes of the antiferromagnetic and ferromagnetic
phases, the radius $R$ of the spheres which form the multiply
connected phase and $y=(N_{\mathrm{F}}-N_{\mathrm{A}})/
(N_{\mathrm{F}}+N_{\mathrm{A}})$, where $N_{\mathrm{F}}$ 
($N_{\mathrm{A}}$) are the number of conduction electrons in the 
FM (AF) phase. $y$ gives the ratio between the excess of electrons in 
a given phase and the total number of electrons and varies from $-1$ 
(all the electrons in the AF phase) to $+1$ (all the electrons in
the FM phase).

With these definitions the densities of conduction electrons in each
phase are:

\begin{eqnarray} 
n_{\mathrm{F}} & = & \frac{N_{\mathrm{F}}}{V_{\mathrm{F}}}
=n(1+x)\frac{1+y}{2}\nonumber\\
n_{\mathrm{A}} & = & \frac{N_{\mathrm{A}}}{V_{\mathrm{A}}}
=n\left(\frac{1+x}{x}\right)\frac{1-y}{2}
\end{eqnarray} 
where $n$ is the mean electron density in the crystal.

The trial energy is given by the expression:
\begin{equation} \label{energy}
E_{\mathrm{ps}}=E_{\mathrm{kin}}+E_{\mathrm{sur}}+E_{\mathrm{C}}+
E_{\mathrm{mag}}
\end{equation}
where 
$E_{\mathrm{kin}}$ is the standard bulk kinetic energy of the          %
conduction electrons, $E_{\mathrm{sur}}$ is the surface electron 
energy, i.e. the correction to $E_{\mathrm{kin}}$ that appears due to
the quantization of the electron motion in regions of finite dimensions 
(in fact, this energy takes into account the electron level spatial
quatization in a bounded region in the Born-Oppenheimer approximation),
$E_{\mathrm{C}}$ is the electrostatic energy  due to the inhomogeneous 
electronic density and $E_{\mathrm{mag}}$ is the magnetic energy.
$E_{\mathrm{mag}}$ is decomposed in two parts: the energy associated to 
the s-d exchange interaction, $E_{\mathrm{s-d}}$ and the one associated
to d-d exchange interaction, $E_{\mathrm{d-d}}$.

We calculate now these energies for the phase-separated and the        %
homogeneous states. As discussed in the introduction, we expect two
different topologies of the phase-separated state depending on the
conduction electron concentration. The conduction electron density at
which this change occurs is denoted by $n_T$. Therefore we divide our
calculation for the energy of the phase-separated state in the 
$n_{\mathrm{d}}<n<n_{\mathrm{T}}$ and $n_{\mathrm{T}}<n<n_{\mathrm{u}}$
cases. As we only consider here conduction electron concentrations over
the $n_d<n<n_u$ range and magnetic fields smaller than the saturation
field of the material, the only possible homogeneous state we
have to consider is the AF one. All the energies are calculated per
unit volume of the unit cell and the numerical values are given in eV.

\subsection{Energy of the phase separated state for 
$n_{\mathrm{d}}<n<n_{\mathrm{T}}$}

Over this concentration range the phase-separated state is expected to %
be an AF matrix in which FM spheres are embedded. The total kinetic 
energy per unit volume of the unit cell is:

\begin{equation}
E_{\mathrm{kin}}=\frac{E_{\mathrm{kin}}^{\mathrm{F}}}{1+x}+
\frac{xE_{\mathrm{kin}}^{\mathrm{A}}}{1+x}
\end{equation}
where $E_{\mathrm{kin}}^{\mathrm{F}}$  
$(E_{\mathrm{kin}}^{\mathrm{A}})$ represent the energies of the 
conduction electrons in the FM (AF) phase.

In the ferromagnetic phase all the charge carriers are assumed to be   %
spin-polarized. For $W<<AS$ this condition is met for all the charge 
carrier densities. For $W>>AS$ it is met when $AS>\mu$ where $\mu$
is their Fermi energy. It is also necessary to take into account the
different band bottoms encountered by the charge carriers moving in 
the FM and the AF part of the crystal \cite{Nag00}. Therefore, in the
free electron approximation, the kinetic energy of the conduction 
electrons in the FM phase is:

\begin{equation} 
E_{\mathrm{kin}}^{\mathrm{F}}=-6tn_{\mathrm{F}}+
\frac{3}{5}\frac{(6\pi^{2}n_{\mathrm{F}})^{\frac{2}{3}}}{2m^{*}
a^{2}}n_{\mathrm{F}}
\end{equation}
From now on $t=\frac{1}{2{m}^{*}{a}^{2}}$ and 
$\mu(n)=t(6\pi^{2}n)^{\frac{2}{3}}$. Then:
 
\begin{equation}
E_{\mathrm{kin}}^{\mathrm{F}}=-6tn(1+x)\frac{1+y}{2}+
\frac{3}{5}\mu(n)n(1+x)^{\frac{2}{3}}\left(\frac{1+y}{2}\right)
^{\frac{5}{3}}
\end{equation}
The kinetic energy of the conduction electrons in the                  %
antiferromagnetic part is: 
\begin{eqnarray} 
E_{\mathrm{kin}}^{\mathrm{A}}=\left\{ \begin{array}{rcl}
-6tn\frac{1+x}{x}\frac{1-y}{2}+\frac{3}{5}
2^{-\frac{2}{3}}\mu(n)n(\frac{1+x}{x})^{\frac{2}{3}}
(\frac{1-y}{2})^{\frac{5}{3}} & \mbox{ if} &
W>>AS\\[1.5ex]
-\frac{6tn}{\sqrt{2S+1}}\frac{1+x}{x}\frac{1-y}{2}
+\frac{3}{5}\frac{\mu(n)n}{\sqrt{2S+1}}(\frac{1+x}{x})^{\frac{2}{3}}
(\frac{1-y}{2})^{\frac{5}{3}} & \mbox{ if} &
W<<AS
\end{array} \right.
\end{eqnarray}

The surface energy, $E_{\mathrm{sur}}$, appears due to the fact that   %
the conduction electrons  are confined in a finite volume. For each
phase, it is proportional to $\frac{S}{V_{i}}$, being $S$ the surface
between both phases and $V_{i}$ the volume of each phase, FM, 
$V_{\mathrm{F}}$ or AF, $V_{\mathrm{A}}$ respectively.
We must point out that actually no surface energy independent of the 
conduction electron density can appear, because this term turns out
meaningless in the case of the conduction electron density equals to 
zero \cite{Nag83}, where phase separation does not exist. 
The contribution to the surface energy due to magnetic interactions is
also zero in the first-neighbour approximation considered by us. 

\begin{equation}
E_{\mathrm{sur}}=\frac{E_{\mathrm{sur}}^{\mathrm{F}}}{1+x}+
\frac{xE_{\mathrm{sur}}^{\mathrm{A}}}{1+x}
\end{equation}
where $E_{\mathrm{sur}}^{\mathrm{F}}$ $(E_{\mathrm{sur}}^{\mathrm{A}})$
represent the surface energies of the conduction electrons in the FM 
(AF) phase.

Taking into account the results of \cite{Bal70}, we have in the FM part%
:
\begin{equation}
E_{\mathrm{sur}}^{\mathrm{F}}=\frac{5}{16}\left(\frac{\pi}{6}\right)
^{\frac{1}{3}} \frac{S}{V_{\mathrm{F}}}\frac{E_{\mathrm{kin}}
^{\mathrm{F}}}{n_{\mathrm{F}}^\frac{1}{3}}
\end{equation}
Using the fact that $\frac{S}{V_{\mathrm{F}}} =\frac{3}{R}$ being $R$ 
(in $a$ units) the radius of the ferromagnetic spheres containing the
conduction electrons, we obtain:

\begin{equation}\label{ EsupFM}
E_{\mathrm{sur}}^{\mathrm{F}}=\frac{9}{16}\left(\frac{\pi}{6}\right)
^{\frac{1}{3}} \frac{1}{R}\mu(n)n^{\frac{2}{3}}(1+x)^{\frac{1}{3}}
\left(\frac{1+y}{2}\right)^{\frac{4}{3}}
\end{equation}	 
For the AF part and making the obvious substitution in equation        %
(\ref{ EsupFM}), we have:

\begin{eqnarray} 
E_{sup}^{A}=\left\{ \begin{array}{r@{\quad \mbox{ if} \quad}l}
\frac{9}{16}\left(\frac{\pi}{6}\right)^{\frac{1}{3}}
\frac{1}{R}2^{-\frac{2}{3}}\mu(n)n^{\frac{2}{3}}
(\frac{1+x}{x^{4}})^{\frac{1}{3}}(\frac{1-y}{2})^{\frac{4}{3}}
& W>>AS\\[1.5ex]
\frac{9}{16}\left(\frac{\pi}{6}\right)^{\frac{1}{3}}
\frac{1}{R}\frac{\mu(n)n^{\frac{2}{3}}}{\sqrt{2S+1}}
(\frac{1+x}{x^{4}})^{\frac{1}{3}}(\frac{1-y}{2})^{\frac{4}{3}}
& W<<AS
\end{array} \right.
\end{eqnarray}

The Coulomb energy $E_{\mathrm{C}}$ is calculated using the jellium    %
model for the ionized impurities. The crystal is separated into the
Wigner cells, i.e. into spheres enveloping the FM inclusions drawn so 
as to make the total charge inside the sphere vanish. This method
provides a good approximation to the electrostatic energy  for small FM
volumes, i.e. in the limit $x>>1$. Moreover it supposes to admit that
the crystal is isotropic and homogeneous regarding the spatial 
distribution of these spheres. So we cannot describe the fractal nature 
of the boundary that separes phases. In this approximation: 

\begin{equation}
E_{\mathrm{C}}=\frac{2\pi}{5\epsilon_{r}}\frac{e^{2}n^{2} R^{2}}
{x^{2}a} \left(\frac{1-y}{2}-x\frac{1+y}{2}\right)^{2}
(2+3x-3x^{\frac{1}{3}}(1+x)^{\frac{2}{3}})
\end{equation} 
 
As was said above, the magnetic energy  has two contributions. One part%
 correspond to the antiferromagnetic interaction between the ionic 
d-shells, $E_\mathrm{{d-d}}$. We estimate it within the mean field 
approximation. Because our calculation is constrained to $T=0$, we only
have to take into account the ground state of the system. For the 
antiferromagnetic phase we have:

\begin{equation}
E_{\mathrm{d-d}}^{\mathrm{A}}=-\frac{|J|S}{2}
\end{equation} 
where $|J|=-\frac{zIS}{2}$.
For the ferromagnetic phase: 

\begin{equation}
E_{\mathrm{d-d}}^{\mathrm{F}}=\frac{|J|S}{2}
\end{equation} 
Therefore the total energy $E_{\mathrm{d-d}}$ is:

\begin{equation}
E_{\mathrm{d-d}}=\frac{xE_{\mathrm{d-d}}^{\mathrm{A}}}{1+x}+
\frac{E_{\mathrm{d-d}}^{\mathrm{F}}}{1+x}=\frac{|J|S}{2}\frac{1-x}{1+x}
\end{equation}
We assume that the AF part is in an collinear AF state. A canted AF    %
state would be also possible. In the case a fourth parameter, namely
the canting angle, must be added to the minimization procedure. Its
value depends only on the conduction electron density, which in the 
AF part of the phase-separated state depends on the parameters $x,y$.
We ruled out this possibility in the view of the results in 
reference \cite{Nag98} where it is shown that the canted state is 
unstable against charge-carrier density fluctuations and it is never
realized.
The second part corresponds to the energy due to the interaction       %
between the conduction electrons and magnetic ions, $E_{\mathrm{s-d}}$.
Again we have:

\begin{equation} 
E_{\mathrm{s-d}}=\frac{xE_{\mathrm{s-d}}^{\mathrm{A}}}{1+x}
+\frac{E_{\mathrm{s-d}}^{\mathrm{F}}}{1+x}
\end{equation}
where in the FM part:

\begin{equation} 
E_{\mathrm{s-d}}^{\mathrm{F}}=-\frac{AS}{2}n_{\mathrm{F}}=
-\frac{AS}{2}n(1+x)\frac{1+y}{2}
\end{equation}
And in the AF part:

\begin{eqnarray} 
E_{\mathrm{s-d}}^{\mathrm{A}}=
\left\{\begin{array}{r@{\quad \mbox{ if} \quad}l}
0 & W>>AS\\[1.5ex]
-\frac{AS}{2}n\frac{1+x}{x}\frac{1-y}{2} & W<<AS
\end{array}\right.
\end{eqnarray} 
Notice that for $W<<AS$, $A$ does not appears in the calculation, it   %
is only an additive constant. 

\subsection{Energy of the phase-separated state 
for $n_{\mathrm{T}}<n<n_\mathrm{u}$}

In the case of that the density of the conduction electrons is higher  %
than the percolation density $n_{\mathrm{T}}$, a phase transition is 
expected with a change in the sample topology. The sample changes from
being a AF matrix in which FM spheres are embedded to have a simply 
connected FM part with AF spheres in. To take this point into account
we have to modify surface and Coulomb energies as follows:

\begin{eqnarray} 
E_{\mathrm{C}} & = & \frac{2\pi}{5\epsilon_{r}}\frac{e^{2}n^2R^{2}x}{a}
\left(\frac{1+y}{2x}-\frac{1-y}{2}\right)^{2}
(2+3x-3x^{\frac{1}{3}}(1+x)^{\frac{2}{3}})\nonumber\\
E_{\mathrm{sur}}^{\mathrm{F}} & = & 
\frac{9}{16}\left(\frac{\pi}{6}\right)^{\frac{1}{3}}
\frac{1}{R}\mu(n)n^{\frac{2}{3}}x(1+x)^{\frac{1}{3}}
\left(\frac{1+y}{2}\right)^{\frac{4}{3}}\\
 E_{\mathrm{sur}}^{A} & = &
\left\{ \begin{array}{r@{\quad \mbox{ if} \quad}l}
\frac{9}{16}\left(\frac{\pi}{6}\right)^{\frac{1}{3}}
\frac{1}{R}2^{-\frac{2}{3}}\mu(n)n^{\frac{2}{3}}
(\frac{1+x}{x})^{\frac{1}{3}}(\frac{1-y}{2})^{\frac{4}{3}}
& W>>AS\\[1.5ex]
\frac{9}{16}\left(\frac{\pi}{6}\right)^{\frac{1}{3}}
\frac{1}{R}\frac{\mu(n)n^{\frac{2}{3}}}{\sqrt{2S+1}}
(\frac{1+x}{x})^{\frac{1}{3}}(\frac{1-y}{2})^{\frac{4}{3}}
& W<<AS
\end{array} \right.\nonumber
\end{eqnarray}
The others energies remain unchanged.

\subsection{Energy of the homogeneous phase}

Over the concentration range $n_{\mathrm{d}}<n<n_{\mathrm{u}}$ the     %
homogeneous antiferro\-mag\-netic phase has an energy:

\begin{equation}
E_{\mathrm{hom}}=E_{\mathrm{kin}}^{\mathrm{hom}}
+E_{\mathrm{mag}}^{\mathrm{hom}}
\end{equation} 
where 

\begin{eqnarray}
\begin{array}{ r@{\quad \quad}l}
E_{\mathrm{kin}}^{\mathrm{hom}} 
=-6tn+\frac{3}{5}2^{\frac{-2}{3}}\mu(n)n
&
E_{\mathrm{mag}}^{\mathrm{hom}}
=-\frac{|J|S}{2}
\end{array}
\end{eqnarray}

if $W>>AS$ and
\begin{eqnarray}
\begin{array}{ r@{\quad \quad}l}
E_{\mathrm{kin}}^{\mathrm{hom}}
=-\frac{6tn}{\sqrt{2S+1}}+\frac{3}{5}\frac{\mu(n)n}{\sqrt{2S+1}}
&
E_{\mathrm{mag}}^{\mathrm{hom}}
=-\frac{|J|S}{2}-\frac{ASn}{2}
\end{array}
\end{eqnarray}
if $W<<AS$

\section{Results}

The stationary state of the system is determined from the condition    %
that the total energy of the system equation (\ref{energy}) be a 
minimum with respect to the three variational parameters $x,\,R$ and
$y$. The parameter $R$ only appears in the surface and Coulomb 
energies. Minimizing the sum $Q=E_{\mathrm{sur}}+E_{\mathrm{C}}$ with
respect to $R$ for fixed $x$ and $y$ leads to the following 
expressions:

\begin{eqnarray} \label{dos}
Q_{\mathrm{min}} & = & \alpha\gamma(hfg^{2})^{\frac{1}{3}}\nonumber\\
R_{\mathrm{min}} & = & \alpha'\gamma'
\left(\frac{g(x,y)}{h(x,y)f(x)}\right)^{\frac{1}{3}}
\end{eqnarray}
where:
\begin{eqnarray}
\begin{array}{r@{\quad  \quad}l}
\alpha=3\left(\frac{2\pi}{5}\right)^{\frac{1}{3}}
\left(\frac{9}{32}\right)^{\frac{2}{3}}\left(\frac{\pi}{6}\right)
^{\frac{2}{9}}\approx1.2
&
\gamma=\left(\frac{n^{\frac{1}{3}}e^{2}}{\epsilon_{r}a}
\mu^{2}n\right)^{\frac{1}{3}}\nonumber\\
\alpha'=\left(\frac{\frac{9}{32}(\frac{\pi}{6})^{\frac{1}{6}}}
{\frac{2\pi}{5}}\right)^{\frac{1}{3}}\approx0.59
&
\gamma'=\left(\frac{\mu(n)n^{-\frac{4}{3}}}{\frac{e^{2}}
{\epsilon_{r}a}}\right)^{\frac{1}{3}}
\end{array}
\end{eqnarray} 
We have used the following definitions:

\begin{eqnarray} 
f(x) & = & 2x+3-3(1+x)^{\frac{2}{3}}\nonumber\\
g(x,y) & = &\left(\frac{1+y}{2}\right)^\frac{4}{3}(1+x)^\frac{1}{3}
+\chi\left(\frac{1-y}{2}\right)^\frac{4}{3}
\left(\frac{1+x}{x^{4}}\right)^\frac{1}{3}\nonumber\\
h(x,y) & = & \frac{1}{x^{2}}
\left(\frac{1-y}{2}-x\frac{1+y}{2}\right)^{2}
\end{eqnarray} 
if $n_\mathrm{d}<n<n_{\mathrm{T}}$. 

In the case of a multiply connected AF phase, i.e. for                 %
$n_{\mathrm{T}}<n<n_\mathrm{u}$, the latter expressions are modified 
according to:

\begin{eqnarray} 
f(x) & = & 2+3x-3x^{\frac{1}{3}}(1+x)^{\frac{2}{3}}\nonumber\\
g(x,y) & = &\left(\frac{1+y}{2}\right)^\frac{4}{3}x(1+x)^\frac{1}{3}+
\chi\left(\frac{1-y}{2}\right)^\frac{4}{3}
\left(\frac{1+x}{x}\right)^\frac{1}{3}\nonumber\\
h(x,y) & = & x\left(\frac{1+y}{2x}-\frac{1-y}{2}\right)^{2}
\end{eqnarray} 
In both cases we have introduced the parameter $\chi$ equal to 
$2^{-\frac{2}{3}}$ if $W>>AS$ and to $(2S+1)^{-\frac{1}{2}}$ if
$W<<AS$.

After substituting equation (\ref{dos}) into equation (\ref{energy}) 
minimization with respect to $x$ and $y$ is carried out numerically. 

The percolation density $n_{\mathrm{T}}$ is calculated by solving for  %
the value of $n$ at which both separate phases are minimal as a 
function of $x$ and $y$ and have the same energy. For $n_{\mathrm{d}}$ 
and $n_{\mathrm{u}}$, the requirements are that the separate state have
a minimum as a function of $x$, and its energy being equal to that of
the corresponding homogeneous state \footnote{Our values for $n_T$ are
slightly different from those in \cite{Nag00}, calculated imposing
$x=y=1$.}.

\begin{figure}[ht]
\begin{center}
\epsfig{file=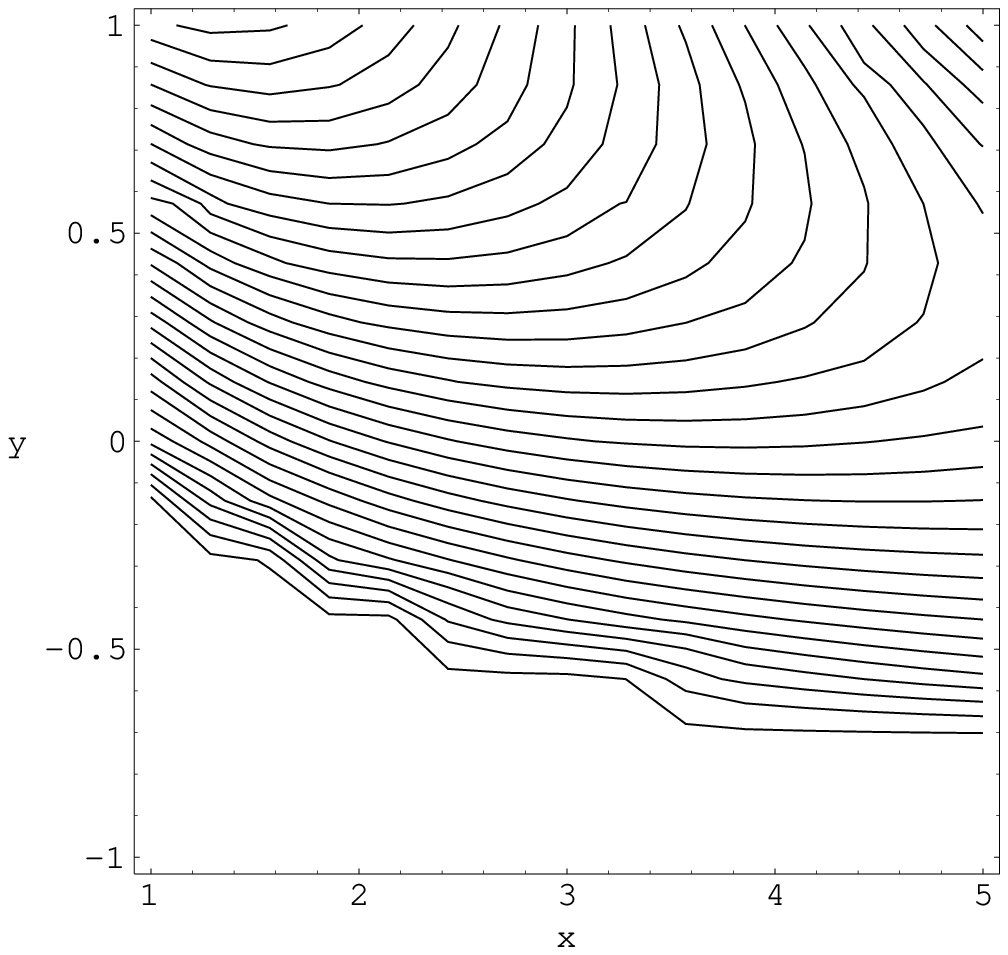,height=7.0 cm}
\end{center}
\caption{Countour plot of $\Delta E$ as a funtion of $x,y$ for 
$W>>AS$ and $n=0.9\,n_{\mathrm{T}}$. 
Lines in the upper part correspond to more
negative $\Delta E$.}\label{fig1}

\begin{center}
\epsfig{file=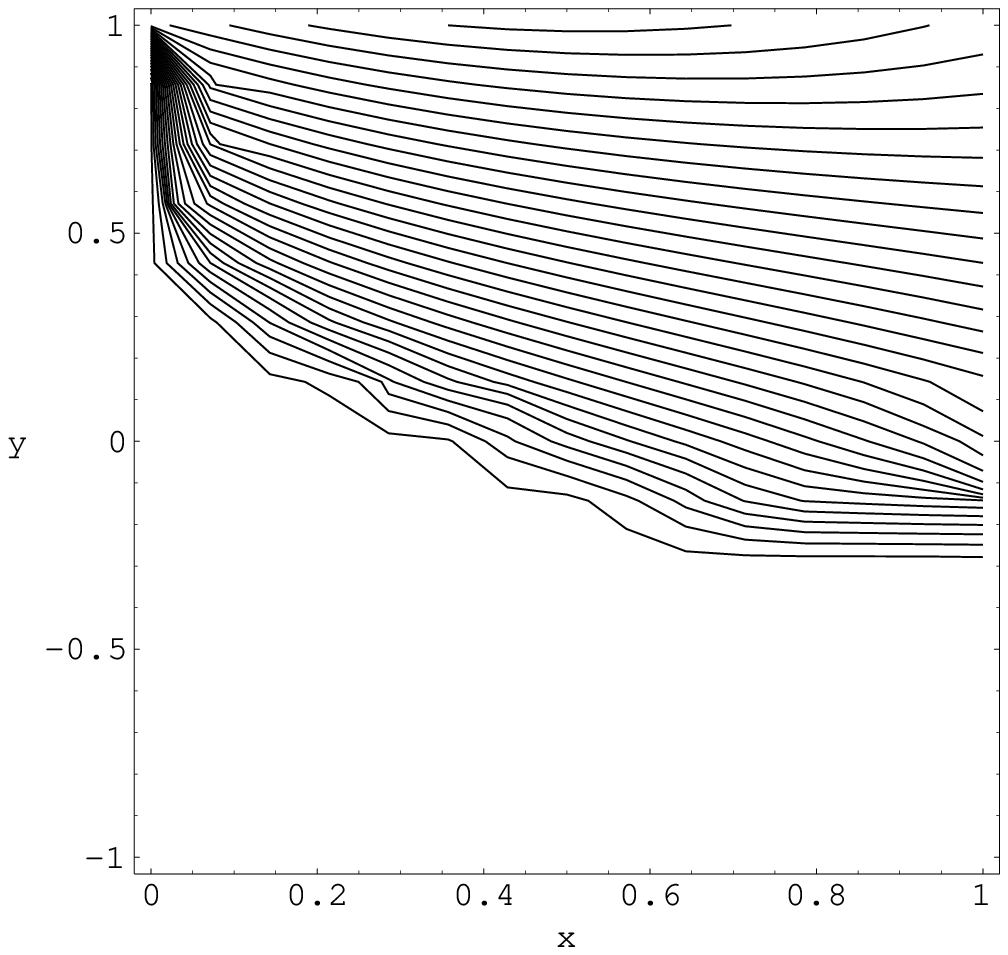,height=7.0 cm}
\end{center}
\caption{Countour plot of $\Delta E$ as a funtion of $x,y$ for 
$W>>AS$ and $n=1.3\,n_{\mathrm{T}}$
Lines in the upper part correspond to more
negative $\Delta E$.}\label{fig2}
\end{figure}

The values of the parameters that characterize wide-band               %
semiconductors are chosen the same as in reference
\cite{Nag90} (they correspond to rare-earth compounds of the $EuTe$ 
type): $S=\frac{7}{2}$, $|J|S=10^{-3}$ eV 
(that implies $T_{\mathrm{N}}=5$ K, and the field at which both 
sublattices collapse  is $98.7$ kOe), $AS=1$ eV,
$\epsilon_{r}=20$, $a^{-3}=4\cdot10^{22}$ cm$^{-3}$, and the electron 
effective mass is equal to free electron  mass, this implies  $W=4$ eV.
The value obtained for  $n_{\mathrm{T}}=1.10\cdot10^{20}$ cm$^{-3}$. 
The value for  $n_{\mathrm{u}}$, at which the conductive
phase-separated state becomes  unstable at $H=T=0$ is  
$1.79\cdot10^{20}$ cm$^{-3}$. In
this work we consider the following electron concentrations
$0.9\,n_{\mathrm{T}}$ and $1.3\,n_{\mathrm{T}}$, as an example of both 
topologies. To study the difference in energies between the energy of 
the homogeneous antiferromagnetic phase $E_{\mathrm{hom}}$ and the 
energy of the phase-separated state $E_{\mathrm{ps}}$ for both electron
concentrations we show in figures \ref{fig1} and \ref{fig2}
a contour plot of the energy difference  
$\Delta E=\frac{E_{\mathrm{ps}}-E_{\mathrm{hom}}-|E_{\mathrm{ps}}
-E_{\mathrm{hom}}|}{2n}$ as a function of $x$ and $y$. Notice that 
$\Delta E=0$ if $E_{\mathrm{ps}}\geq E_{\mathrm{hom}}$. The lines on 
the diagram are  lines of constant $\Delta E$. 
The regions with no lines correspond to $\Delta E=0$. The lines in the
upper part of the figures correspond to more negative energy 
differences.
It can be observed that as $y$
increases $\Delta E$ becomes more negative. In both cases, the minimum 
value of $\Delta E$ is obtained for $y=1$, i.e. $N_{\mathrm{A}}=0$.
For the values of $n$ we show here, the minima are 
$\frac{E_{\mathrm{ps}}-E_{\mathrm{hom}}}{n}(x=1.3662,y=1)=-0.1707$ at  %
$n=0.9\,n_{\mathrm{T}}$ and $\frac{E_{\mathrm{ps}}-E_{\mathrm{hom}}}
{n}(x=0.5169,y=1)=-0.2095$ at $n=1.3\,n_{\mathrm{T}}$.

\begin{figure}[ht]
\begin{center}
\epsfig{file=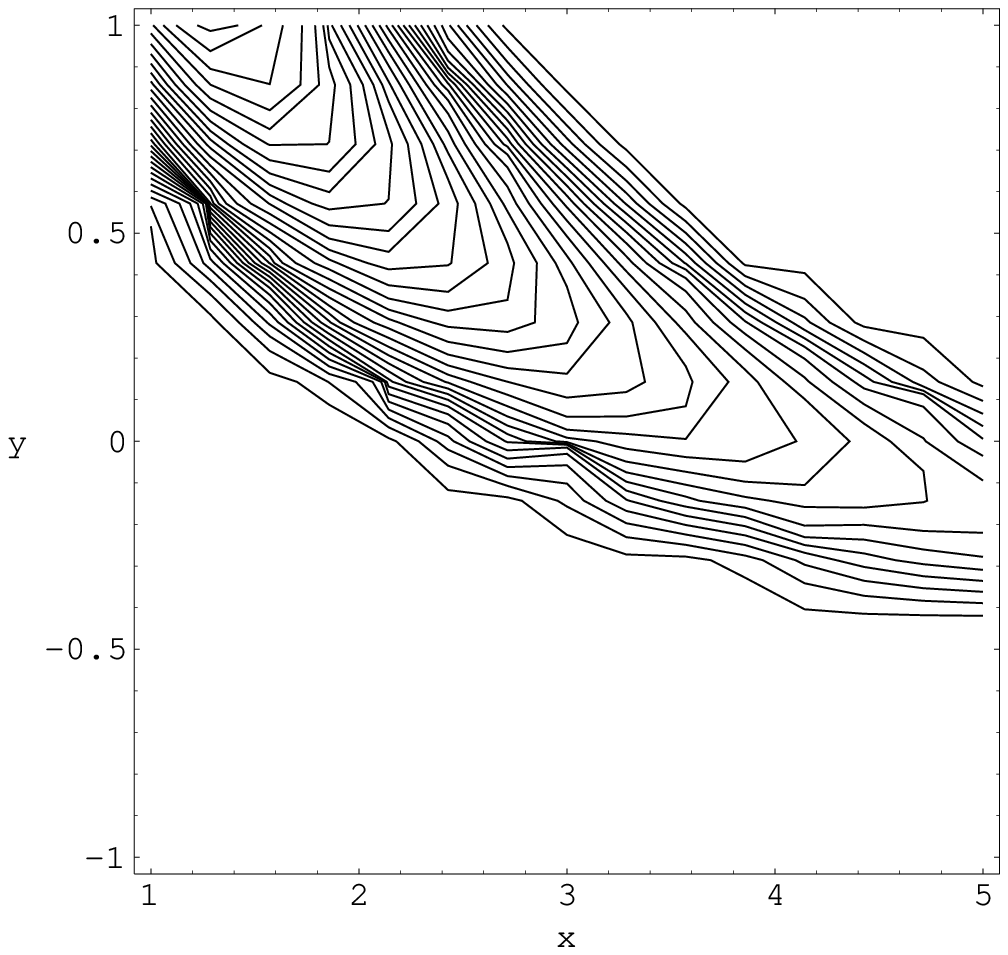,height=7.0 cm}
\end{center}
\caption{Countour plot of $\Delta E$ as a funtion of $x,y$ for 
$W<<AS$ and $n=0.9\,n_{\mathrm{T}}$
Lines in the upper part correspond to more 
negative $\Delta E$.}\label{fig3}

\begin{center}
\epsfig{file=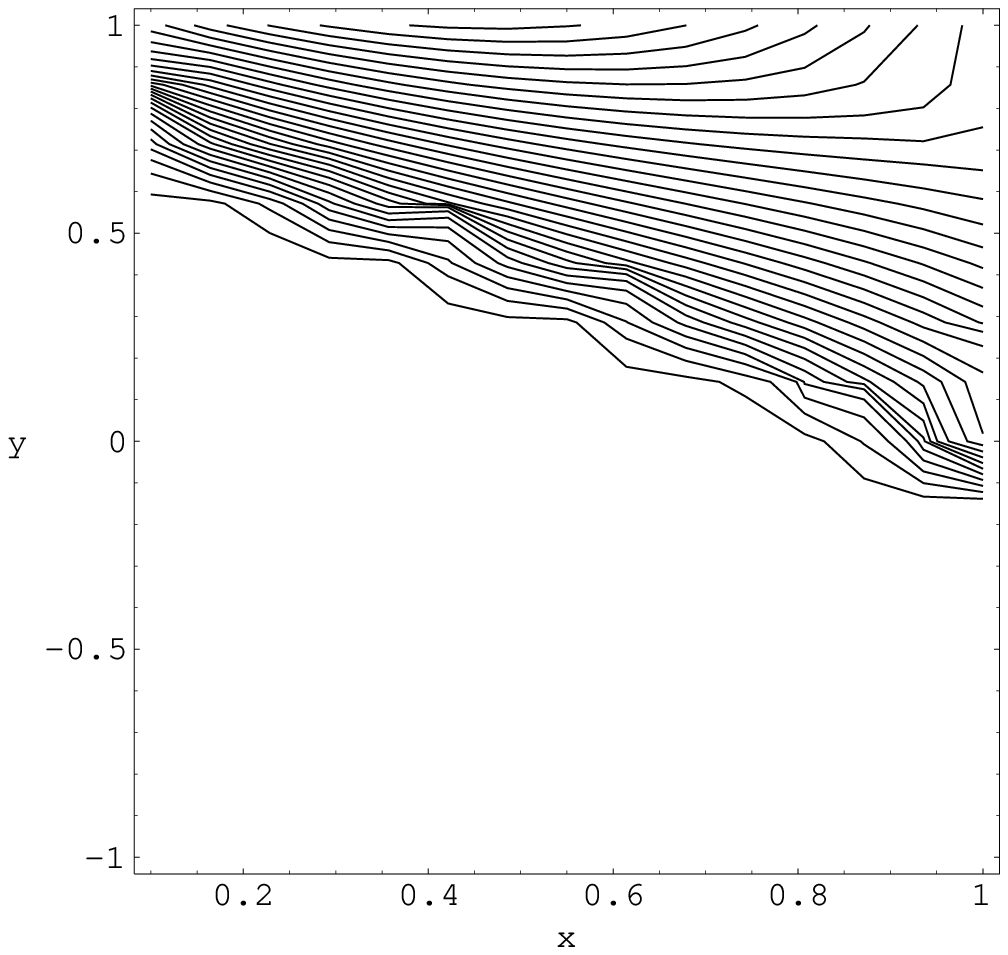,height=7.0 cm}
\end{center}
\caption{Countour plot of $\Delta E$ as a funtion of $x,y$ for 
$W<<AS$ and $n=1.3\,n_{\mathrm{T}}$
Lines in the upper part correspond to more
negative $\Delta E$.}\label{fig4}
\end{figure}

The values of parameters that characterize ``double-exchange''         %
semiconductors  are the following (typical for lanthanium manganites, 
for example): $S=2$, $D=-zIS^{2}=10^{-2} \mathrm{eV}$ (that implies 
$T_{\mathrm{N}}=58$ K, and the field at which the two sublattices
collapse at $T=0$ is $1.73$ MOe), $\epsilon_{r}=5$, 
$a^{-3}=4\cdot10^{22}$ cm$^{-3}$,  the effective electron mass equal to
the free electron mass, that implies $W=4$ eV. $A$ is regarded as high 
enough to meet $W<<AS$, although its precise value is not needed
for calculation. We obtain $n_{\mathrm{T}}=5.93\cdot 10^{20}$ cm$^{-3}$
and $n_{\mathrm{u}}=9.46\cdot10^{20}$ cm$^{-3}$. As before we consider 
the following electron concentrations $0.9\,n_{\mathrm{T}}$ and 
$1.3\,n_{\mathrm{T}}$. The energy minima are
$\frac{E_{\mathrm{ps}}-E_{\mathrm{hom}}}{n}(x=1.3159,y=1)=-0.04454$ 
at $n=0.9\,n_{\mathrm{T}}$ and
$\frac{E_{\mathrm{ps}}-E_{\mathrm{hom}}}{n}(x=0.4681,y=1)=-0.1162$
at $n=1.3\,n_{\mathrm{T}}$. 
Contour plots are shown in figures \ref{fig3} 
and \ref{fig4}. 
In both cases minima are found in $y=1$, i.e. all the conduction 
electrons are in the AF part of the phase-separated states. 

Analogous results are obtained for any value of the concentration of   %
conduction electrons over the range 
$n_{\mathrm{d}}<n<n_{\mathrm{u}}$. Therefore, we have
shown that for both kinds of compounds and over the above mentioned
range of concentrations, the ground state corresponds to a 
phase-separated state with FM and AF parts, with absence of electrons 
in the antiferromagnetic part.

\section{Acknowlegments}
The authors are deeply indebted to Prof. E. L. Nagaev for proposing 
the problem and his encouragement and help during its realization.
Unfortunately, Prof. Nagaev died recently. This article is dedicated to
his memory.

\end{document}